\documentstyle[preprint,aps,prl]{revtex}

\begin{document}

\draft
%\preprint{}

\title{Room Temperature Molecular Single-Electron Transistor}

\author{E. S. Soldatov$^1$, V. V. Khanin$^1$, A. S. Trifonov$^1$, 
S. P. Gubin$^2$, V. V. Kolesov$^3$, D. E. Presnov$^1$, 
S. A. Iakovenko$^1$, G. B. Khomutov$^1$, and A. N. Korotkov$^4$.}

\address{
$^1$Department of Physics, Moscow State University, 
        Moscow 119899, Russia, \\
$^2$Institute of General and Inorganic Chemistry, Moscow 117907, Russia, 
\\
$^3$Institute of Radioengineering and Electronics, Moscow 103907, 
Russia,\\
and 
$^4$Nuclear Physics Institute, Moscow State University, 
        Moscow 119899, Russia}

\date{\today}

\maketitle

\begin{abstract}
        A room temperature single-electron transistor based on
the single cluster molecule has been demonstrated for the first
time. Scanning tunneling microscope tip was used to study the transport
via single carboran cluster molecule 
1,7-(CH$_3$)$_2$-1,2-C$_2$B$_{10}$H$_9$Tl(OCOCF$_3$)$_2$ incorporated
into the Langmuir-Blodgett monolayer of the stearic acid.
DC {\it I-V} curves at 300 K showed well pronounced Coulomb staircase, 
and the current could be controlled by a lithographically fabricated
gold gate electrode.

\end{abstract}


\narrowtext

\vspace{1ex}

        A considerable interest was attracted to the effects of 
correlated single-electron tunneling \cite{Av-Likh} during the last 
decade.
The most of experiments with lithographically fabricated structures
were done at temperatures below 1 K. To increase the operating
temperature $T$ it is necessary to reduce the characteristic size $d$ 
of the structure in order to decrease the typical capacitance $C$.
Operation at $T=300$ K requires $C\alt 10^{-18}$ F corresponding
to $d\alt 3$ nm, quite a difficult task. 
        Several different fabrication techniques were studied to 
obtain single-electron effects at relatively high temperatures
(for a recent review see, e.g., Ref.\ \cite{Kor-rev}). Let us
mention, for example, room temperature memory effects \cite{Yano},
a single-electron transistor with visible charging effects up to 300 K
\cite{Takahashi} using silicon-based structures, a gated system
of metal grains \cite{Ahmed} operating at 77 K, and room-temperature
single-electron transistor fabricated by nanooxidation process 
\cite{Matsumoto}. 

        Another technique in which the small capacitances are
easily obtained is based
on scanning tunneling microscopy (STM). The simplest
single-electron circuit consisting of two tunnel junctions
in series can be implemented using the STM tip, small
conducting particle, and the substrate. The single-electron
charging  survives up to
room temperature for sufficiently small metal particles
\cite{Schonenberger,Dorogi}, and it can be even stronger when the
tunneling via single molecules is studied 
\cite{Nejoh,Fischer,Zubilov,Dubois}.
        The drawback of this technique was the absence of the gate
electrode to control the transport. In the present letter we report
the first demonstration of a room temperature
molecular single-electron transistor (the schematic is shown in Fig.\
\ref{scheme}) with a metallic gate 
controlling the electron tunneling from
the STM tip to the substrate via single carboran cluster molecule 
(see also Ref.\ \cite{Soldatov}).

        The molecules with the ``core'' consisting of the cluster
of metal atoms and surrounded by ligands are being extensively 
studied by both chemical and physical methods during last few 
decades (see, e.g., Ref.\ \cite{Kreibig} and references therein). 
The important property of these molecules is
their ability to accept as well as to release several electrons 
without the structural change. The similar quasimetallic 
property can be found in some cluster molecules with nonmetallic 
core \cite{Kreibig}. In our experiment the Tl-derivative 
1,7-(CH$_3$)$_2$-1,2-C$_2$B$_{10}$H$_9$Tl(OCOCF$_3$)$_2$ 
of carboran (which has the nonmetallic core) was used 
(for brevity we call this cluster molecule simply as carboran).

        Langmuir-Blodgett (LB) monolayers of stearic acid with 
incorporated cluster molecules were deposited 
on the pyrolytic graphite (HOPG) substrate with pre-formed 
gate electrode. This electrode was fabricated by the conventional 
electron lithography and presented a system of thin 
(50 nm) and narrow (400 nm) gold strips separated by 400-nm
gaps from each other (Fig.\ \ref{gate}). All strips were
electrically connected and separated from the substrate by a
50 nm thick insulator (Al$_2$O$_3$) -- see Fig.\ \ref{scheme}. 
 
        The LB monolayer was deposited using a computerized 
version on the 
conventional teflon trough by the technique described in more 
detail in Refs.\ \cite{Zubilov} and \cite{Iakovenko} (see also
Ref.\ \cite{Gubin-new}).
The small volume (about 0.1 ml) of the mixture (1:36) 
of the carboran clusters and 
the stearic acid dissolved in tetrahydrofuran (total molecular 
concentration 0.001 M) was put on the surface of MilliQ-purified 
water. After the complete evaporation of tetrahydrofuran 
the remaining monolayer was
compressed at a speed about 5 \AA$^2$/molecule/min. 
The LB monolayer was transferred from the trough onto the substrate 
with gate electrode by Schaefer's method \cite{Shaefer}.

        Electron transport through the film was studied at 
room temperature by NanoScope STM having an atomic resolution. 
First the films were imaged at the typical tip bias voltage of 0.1 V 
and the tunneling current of 0.5 nA. The images were stable and 
reproducible. The carboran clusters were seen as elevated oval shape
objects with the longer size about 2 nm and the shorter
size crudely twice less. The average distance between clusters 
was about 20 nm. The size and the shape were not strictly
reproducible, that can be explained, e.g., by the different
orientations of the cluster molecules in the monolayer.

        After finding a  cluster molecule, the STM tip was positioned 
above the cluster and a series of the transport measurements with
disconnected STM feedback loop was carried out (the feedback loop
with voltage of 0.1 V and current of 0.5 nA was 
used between measurements to restore the vertical position of the tip).
Actually the measurements were carried in 49 closely located points 
in the plane, so that the points directly above the cluster as well as
the points aside the cluster were studied (similar technique was used
in Ref.\ \cite{Zubilov}).
        The typical dc {\it I-V} curve in the case of tunneling via
the cluster  molecule is shown in Fig.\ \ref{I-V}. It has a clear 
staircase shape (with voltage period about 130 mV) while there are 
no stairs if the STM tip is far from the cluster.
        Figure \ref{I-V} also shows the result of the differential 
conductance measurement using lock-in technique. Notice that the 
voltage position of the conductance oscillations corresponds well 
to the staircase structure, however, the measured conductance
is somewhat larger than what it could be expected from the numerical 
derivative of the {\it I-V}
curve. The difference can be explained by the slight deviation 
of the tip position between the measurements 
(the vertical drift is about 0.6 \AA/sec and in-plane drift is about
0.2 \AA/sec; the duration of one {\it I-V} curve measurement 
is about 10 msec while one conductance curve measurement takes 
about 3 seconds).
Another reason
is the fact that the curves in Fig.\ \ref{I-V} actually show the 
characteristics averaged over 4 independent sweeps (the individual
curves were typically too noisy).

        The curve 1 in Figure \ref{control} shows a dependence 
of the tunneling 
current $I$ on the gate voltage $V_g$ for the case when the STM tip 
was positioned above the cluster located about 60 nm from the gate 
electrode. The curve is clearly periodic with the period about 0.8 V. 
In contrast, $I-V_g$ curves do not show such an effect when the STM 
tip is positioned above the flat area of surface without clusters
(curve 2).
        Assuming that each period of the oscillations of the curve 1 
corresponds to one additional electron, we can estimate the charge
sensitivity of our molecular single-electron transistor as $10^{-3} e/
\mbox{Hz}^{1/2}$ (the measurement system had a bandwidth of 16 kHz).
        We suppose that mechanical vibrations give the main contribution 
to the total system noise ($\sim$150 pA peak-to-peak in 
Fig.\ \ref{control}).
        The modulation amplitude of the $I-V_g$ curve 
depends on the dc bias voltage $V$. Qualitatively we have found that the
amplitude is maximal for $V$ corresponding to the points between the
steps of {\it I-V} curve while the gate control is almost negligible
when $V$ corresponds to a flat part of the curve.
         We should note that the clearly periodic
control curves for the tunneling via cluster were observed quite rare 
(curve 1 in Fig.\ \ref{control} presents the best result), 
that complicated their quantitative analysis.

        The experimental results qualitatively agree with the
theory of the single-electron transistor \cite{Av-Likh}, however,
several issues are not clear yet. The single electron tunneling
in a molecular system should necessarily be affected 
\cite{Kor-rev,Aver-Kor} by the discreteness of the energy spectrum. 
In this case the electric capacitance becomes a not well-defined
quantity, and the Coulomb energy converts into the energy of 
ionization and electron affinity. 
However, at the present time there is no independent information 
in the literature about the energy spectrum of the studied system.
On the other hand, the simple ``orthodox'' theory \cite{Av-Likh} 
which assumes the continuous energy spectrum of the electrodes
is proven to work sufficiently well even for nanometer-size 
systems.

        So, as a first approximation it is natural to try the explanation
of the experimental {\it I-V} curve shape using the ``orthodox'' theory.
 Then taking the staircase period of $\Delta V \simeq$130 
mV from Fig.\ \ref{I-V} we can calculate the capacitance of the
junction with larger resistance as $C=e/\Delta V \simeq 1.2\times 
10^{-18}$ F. 
        This value is too large to be explained as the capacitance
of the core of the carboran cluster which is 0.7 nm in  diameter.
STM image of the cluster  molecule is also considerably 
larger (about 2 nm). It is reasonable to assume that the 
effective size (in the sense of electric charging energy) is larger
that the core size \cite{notice}. The sphere of the diameter 
$d=2$ nm has capacitance $C\simeq \varepsilon \times 1.1\times 
10^{-19}$ F. Hence, the experimental capacitance of 
$1.2\times 10^{-18}$ F can be reasonably well explained if we
assume the effective dielectric constant $\varepsilon
\sim 5$ (related to the substrate, stearic acid, and adsorbate)
and take into account the capacitance increase when geometry is 
closer to the plane capacitor. 

        Figure \ref{theory} shows the $I-V$ curves calculated
using the ``orthodox'' theory (curves 1 and 2) to fit the 
experimental result (curve 3). One can see that the agreement
between the theory and the experiment is quite good qualitatively
but not perfect quantitatively.
        The calculations show that to have so well pronounced
Coulomb staircase the temperature should be more than 1.5 times
smaller than it was experimentally.
        Almost horizontal steps obtained in the 
experiment can be explained in the ``orthodox'' theory assuming 
that the tunnel junction which has much higher resistance, has 
also considerably larger capacitance. Though such an assumption 
does not seem to be quite natural, it is the usual explanation 
of the almost flat steps in single-electron experiments using STM 
(see, e.g., Ref.\ \cite{Wilkins}).
        The horizontal steps could be also explained if we assume
the discrete energy spectrum of the central electrode of transistor
\cite{Aver-Kor}, that is quite natural for the molecular system. 

        The gate capacitance $C_g$ calculated from the period of the 
control curve (curve 1 in Fig.\ \ref{control}) 
is about $2\times 10^{-19}$ F.
The ratio $C_g/C \sim 0.15$ is relatively large although the
distance between the cluster and the gate electrode ($\sim$60 nm) is
much larger than the typical distance between the STM tip and the 
cluster molecule.  The possible explanation can be based on the fact that 
the graphite substrate is quite far from being a perfect conductor, and
this considerably reduces the screening of the gate voltage by 
the bias electrodes. The theoretical fitting of the 
modulation amplitude of the control curve 1 shown in Fig.\ \ref{control}
(using the capacitance calculated from Fig.\ 
\ref{I-V}) gives the discrepancy of about 1.5 times in the temperature
similar to the discrepancy in the {\it I-V} curve fitting. 
However, Fig.\ \ref{control} shows the largest current swing obtained,
and the typical control curves do not go beyond the limit 
explainable by the ``orthodox'' theory.

        In conclusion, we have demonstrated the controllable 
single-electron system based on a single cluster molecule. 
The clear Coulomb staircase and the transistor action were
obtained at room temperature. The experimental results are in
a good qualitative agreement with the ``orthodox'' theory of 
the single-electron transistor, however,
there are several quantitative discrepancies. Further
studies of this subject are certainly necessary.

        The work was supported in part by the Russian  
program on the prospective technologies for nanoelectronics (Grant No.\ 
5-029/26/1-3), the Russian program on the physics of 
nanostructures (Grant No.\ 1-091/4) and the Russian 
Foundation for Basic Research (Grant No. 96-03-33766a).

        \begin{figure}
\caption{ Schematic of the single-electron transistor based
on the single cluster molecule. 1 -- HOPG substrate, 2 --
insulating layer (Al$_2$O$_3$), 3 -- gold gate electrode.}
\label{scheme} \end{figure}

        \begin{figure}
\caption{SEM image of the gold gate electrode fabricated
before the monolayer deposition.}
\label{gate} \end{figure}

        \begin{figure}
\caption{The typical {\it I-V} curve and the differential conductance
(as functions of the bias voltage $V$) of the molecular 
single-electron transistor.}
\label{I-V} \end{figure}

        \begin{figure}
\caption{Curve 1 -- the dependence of the current through the molecular 
single-electron transistor on the gate voltage (dc bias point
is shown by arrow in Fig.\ \protect\ref{I-V}). Curve 2 -- similar 
dependence when STM tip is positioned above the stearic acid 
(without cluster).
The bias voltage for both curves is 0.2 V.}
\label{control} \end{figure}

        \begin{figure}
\caption{The $I-V$ curves calculated using the ``orthodox'' theory 
(curves 1 and 2) to fit the experimental result (curve 3). Curves
are shifted vertically for clarity.}
\label{theory} \end{figure}


\begin{references}
\bibitem{notice} We should note that in our earlier study of the
        similar system \cite{Zubilov} (without gate electrode) 
        we did not observe the pronounced Coulomb staircase, but 
        observed considerably larger ($\sim$ 500 mV) Coulomb blockade.
        It is not excluded that additional fabrication step has
        changed the chemical environment of the cluster  molecules.
\end{references}
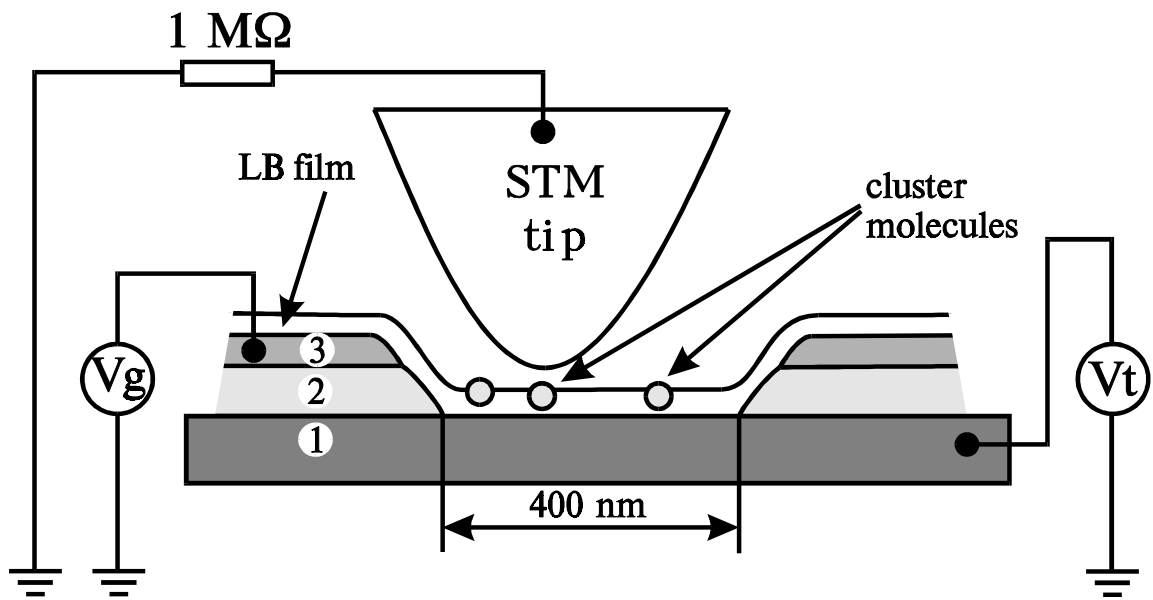
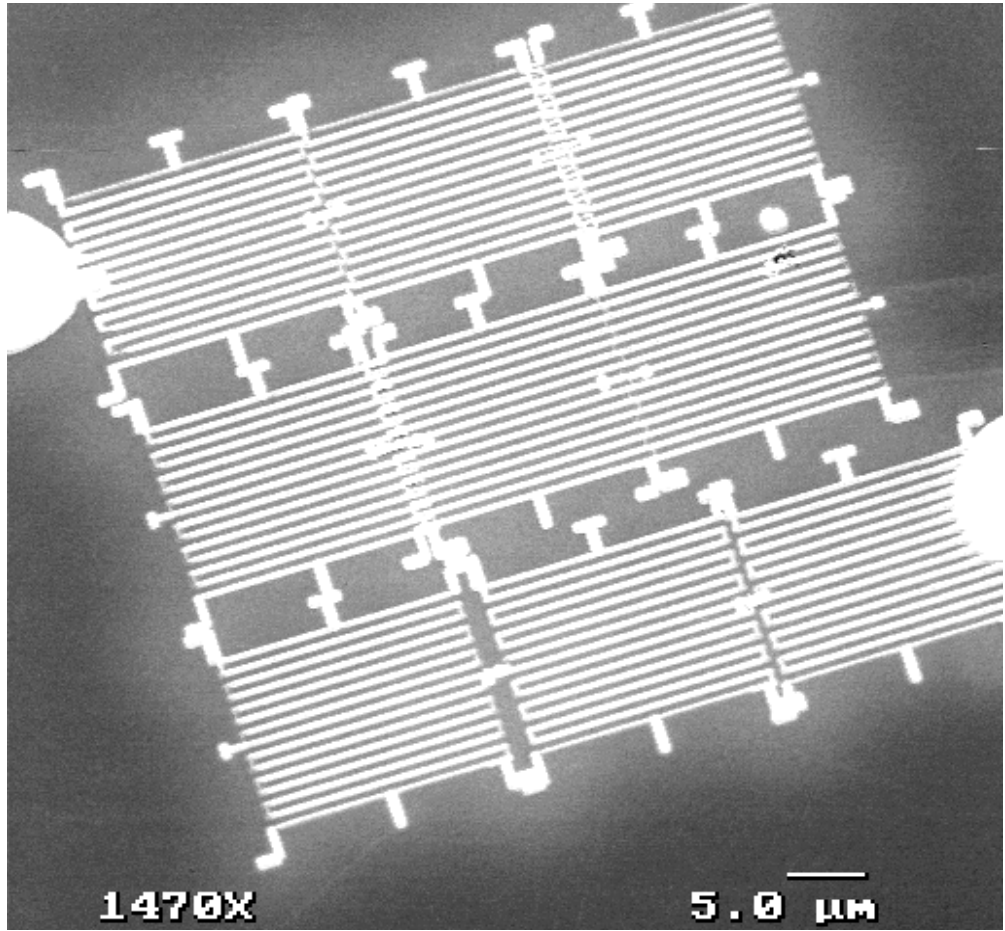
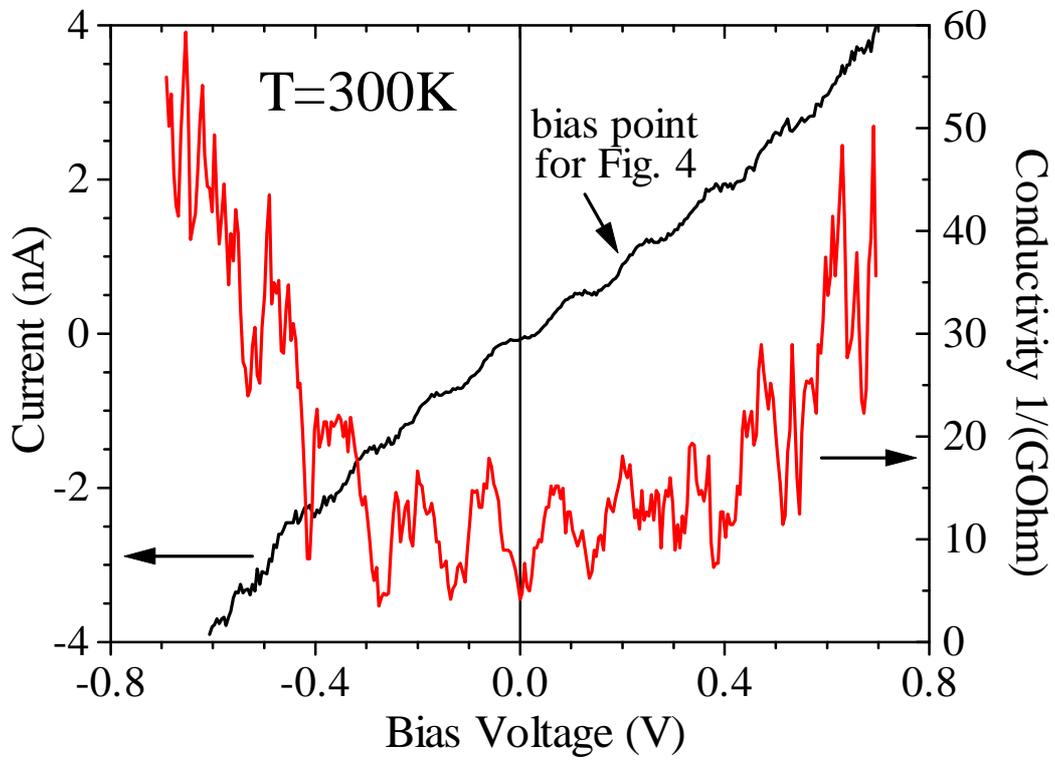
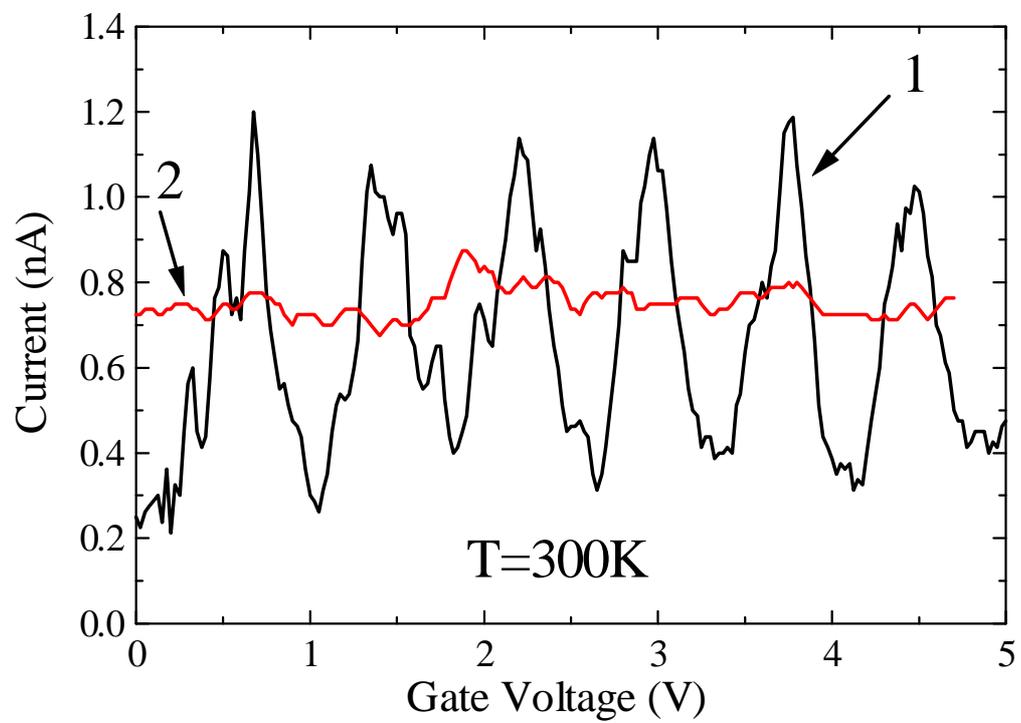
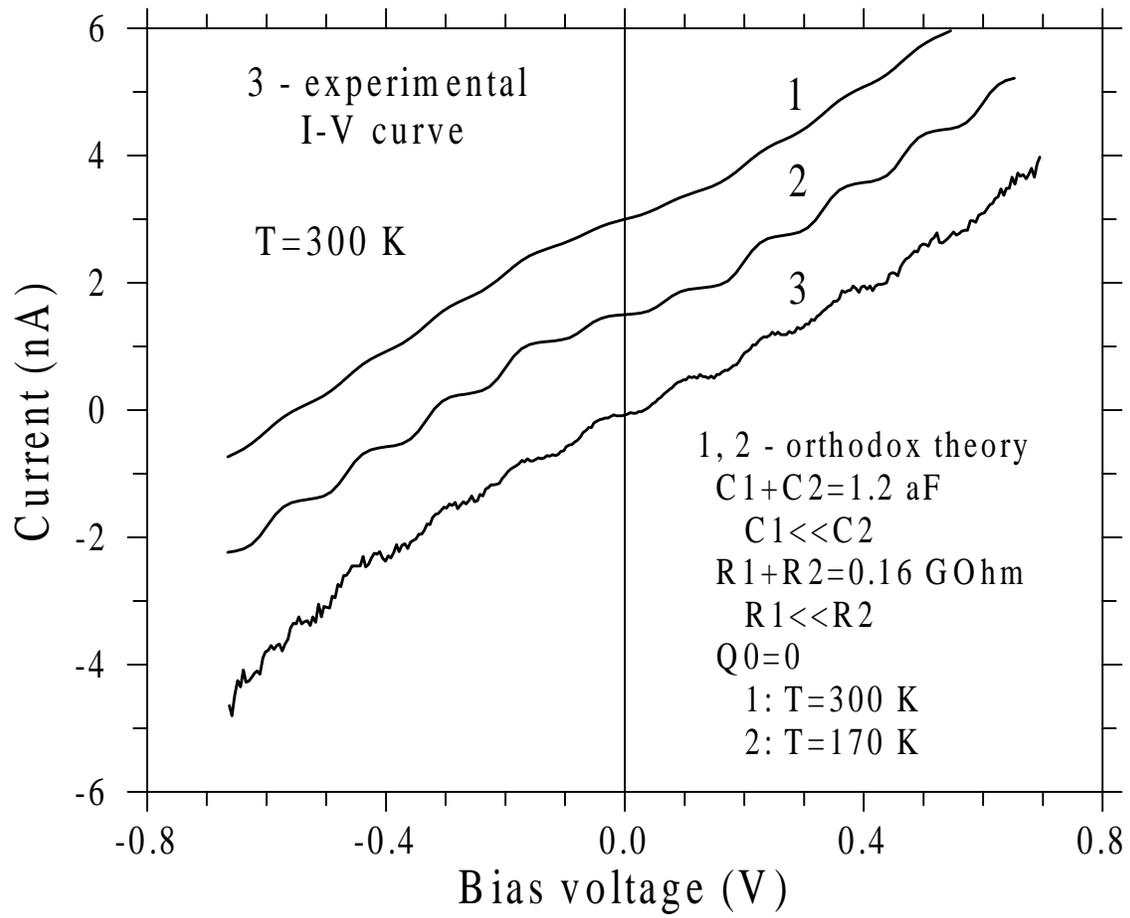
\end{document}